\documentclass{elsart3p}

\usepackage{natbib}
\input epsf
\usepackage{epsfig}
\usepackage{amssymb}
\usepackage{amsmath}
\def\plotone#1{\centering \leavevmode
\epsfxsize=2 \columnwidth \epsfbox{#1}}
\def\plottwo#1#2{\centering \leavevmode
\epsfxsize=\columnwidth \epsfbox{#1} \hfil
\epsfxsize=\columnwidth \epsfbox{#2}}

\begin{document}

\begin{frontmatter}

\title{Difficulties with the QPOs Resonance Model}


\author{Paola Rebusco}

\address{Kavli Institute for Astrophysics and Space Research, MIT, Cambridge, MA 02139, USA}
\thanks{Supported by the Pappalardo Postdoctoral Fellowship in Physics at MIT.}

\hspace{10 mm}

\begin{quote}
\begin{flushright}
\small{"To know that we know what we know,\\ 
and to know that we do not know what we do not know,\\ 
that is true knowledge."\\[2 mm]
\textit{Nicolaus Copernicus}}
\end{flushright}
\end{quote}

\begin{abstract}
% Text of abstract
High frequency quasi-periodic oscillations (HFQPOs) have been detected in microquasars and neutron star systems. The resonance model suggested by Klu\'{z}niak \& Abramowicz (2000,2001) explains twin QPOs as two weakly coupled nonlinear resonant epicyclic modes in the accretion disk. Although this model successfully explains many features of the observed QPOs, it still faces difficulties and shortcomings. Here we summarize the aspects of the theory that remain a puzzle and we briefly discuss likely developments.
\end{abstract}

\begin{keyword}
accretion \sep general relativity \sep X-rays: binaries \sep  nonlinear resonance \sep QPOs 

\end{keyword}

\end{frontmatter}

% main text
\section{Introduction}
\label{sec:intro}

Many galactic black hole and neutron star sources in low-mass X-ray binaries show quasi-periodic variability in their observed X-ray fluxes. Some of the quasi-periodic oscillations (QPOs) are of the order of few hundred Hz (hence the name high frequency) and often come in pairs of twin peaks in the Fourier power spectra \citep[see][for a review]{van00,rem06}. High frequency QPOs draw much interest because their frequencies lie in the range of orbital frequencies few Schwarzschild radii outside the central source. Furthermore for different sources they scale with $1/M$, where $M$ is the mass of the central compact object \citep[][]{mcc06}. These two characteristics make QPOs attractive tools to possibly test General Relativity in strong fields.
Many models have been proposed in order to understand HFQPOs: some of them involve orbital motions \citep[e.g.,][]{klu90,ste98,lam03,sch04}, others consider accretion disk oscillations \citep[e.g.,][]{wag01,kat01,rez03,kat07,tas07}. None is definitive.
Klu\'{z}niak \& Abramowicz (2000,2001) proposed that certain properties of HFQPOs in neutron stars could 
be explained by resonant motions of accreting fluid in strong gravity, 
and predicted that in black hole systems 
two HFQPOs in a frequency ratio of 1:2, 1:3, n:m,  could be detected. 
\citet{abr01} observed that the newly discovered HFQPOs in GRO J 1655-05 \citep{str01} 
were in a $3:2$ ratio, and proposed a more specific model of resonance. 
This idea guided successive studies \citep[e.g.,][]{abr03,reb04,hor04,klu04} aimed at clarifying and expanding the theory. Wlodek Klu\'{z}niak talked about the accomplishments of the model  (see the contribution in the same Proceedings), here we focus on its limits.
% Bibliographic references with the natbib package:
% Parenthetical: \citep{Bai92} produces (Bailyn 1992).
% Textual: \citet{Bai95} produces Bailyn et al. (1995).
% An affix and part of a reference:
%   \citep[e.g.,][Ch. 2]{Bar76}
%   produces (e.g., Barnes et al. 1976, Ch. 2).

\begin{figure*}
\plotone{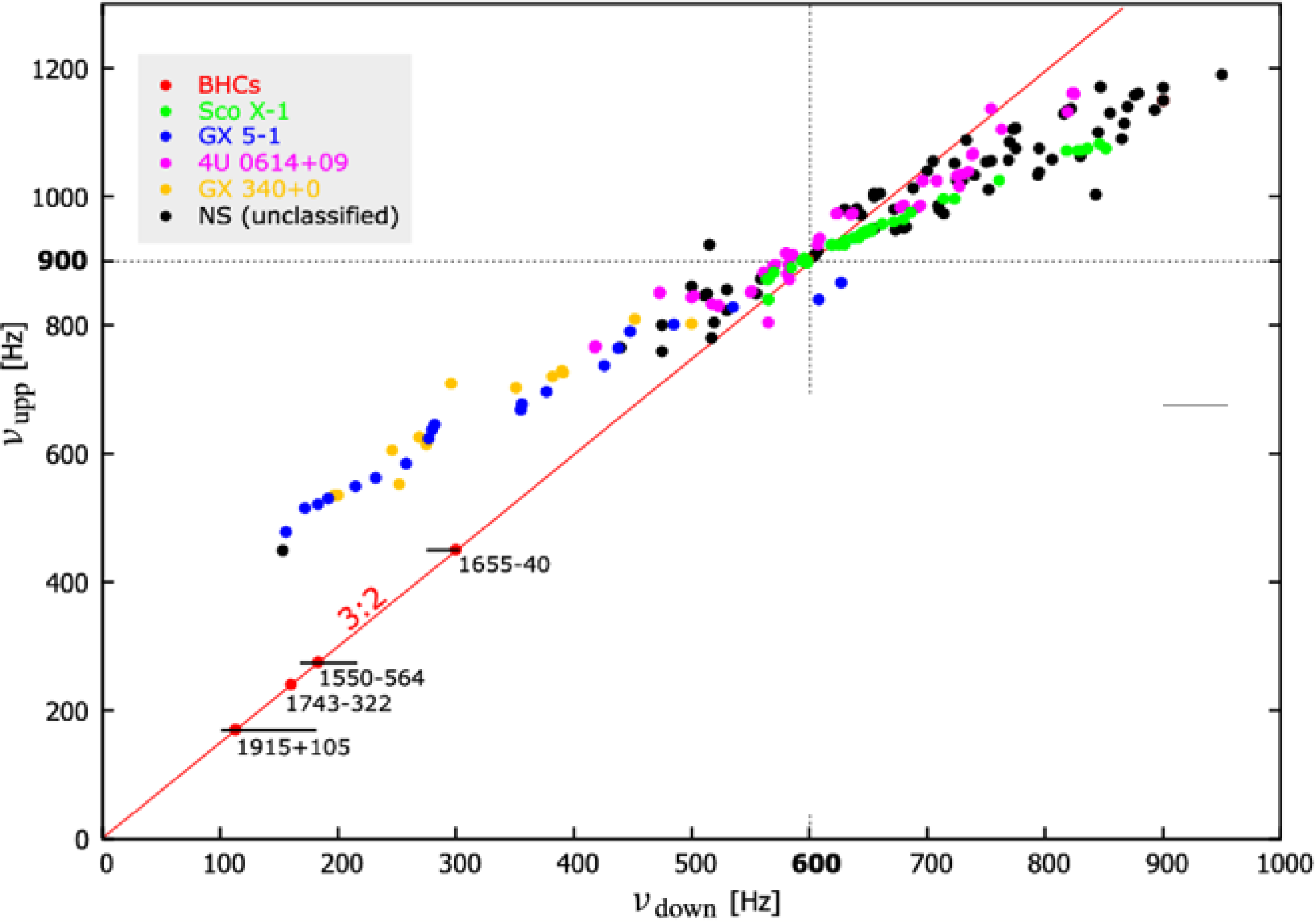}
\caption{Lower ($\nu_{down}$) versus upper ($\nu_{upp}$) frequencies in observed twin peak HFQPOs. The four microquasars lie on the lower left part of the plot, on the line with slope $3:2$ . Neutron star systems (all the others) cross the $3:2$ line, but at different times of observation may be in a different position and ratio (courtesy of Michal Bursa). }
\label{fig:bursa}
\end{figure*}

\section{High Frequency QPOs and Nonlinear Resonances}
\label{sec:QPOs}

The fundamental features of the nonlinear resonance model are that:\\

\begin{itemize}
\item HFQPOs arise from a nonlinear resonance in accretion disks in General Relativity.\\ 
\item The frequencies of nonlinear oscillations depend on the amplitudes of the same oscillations. As a consequence the observed frequencies ($\nu^*$) may differ from the
 eigenfrequencies ($\nu_0$) and can vary in time ($\nu^*=\nu_0+\Delta\nu(t)$). We refer to this well-known nonlinear phenomenon as frequency correction.\\
\item The two frequencies of resonant modes are approximately in the ratio of small integers, most likely in the ratio $3:2$.
\end{itemize}

\section{Difficulties with the Nonlinear Resonance Model}
\label{sec:difficult}

All models of QPOs are essentially dynamical models, that miss any emission mechanisms and any connections to the spectral states of the sources. The only seminal steps in this direction have been done by \citet{bur04} and \citet{sch04}, who considered general relativistic ray-tracing from an oscillating slim torus and an orbiting hot-spot respectively. 
In what follows we concentrate on the difficulties in the pure dynamics, since the radiation problem still has to be addressed by all the models. \\
HFQPOs from black holes are stable (for many years), while neutron star QPOs are variable and present a rich phenomenology: consequently the two classes of systems will be treated separately.

\subsection{Difficulties in Black Hole Systems}
\label{subsec:bh}

\begin{figure*}
\plottwo{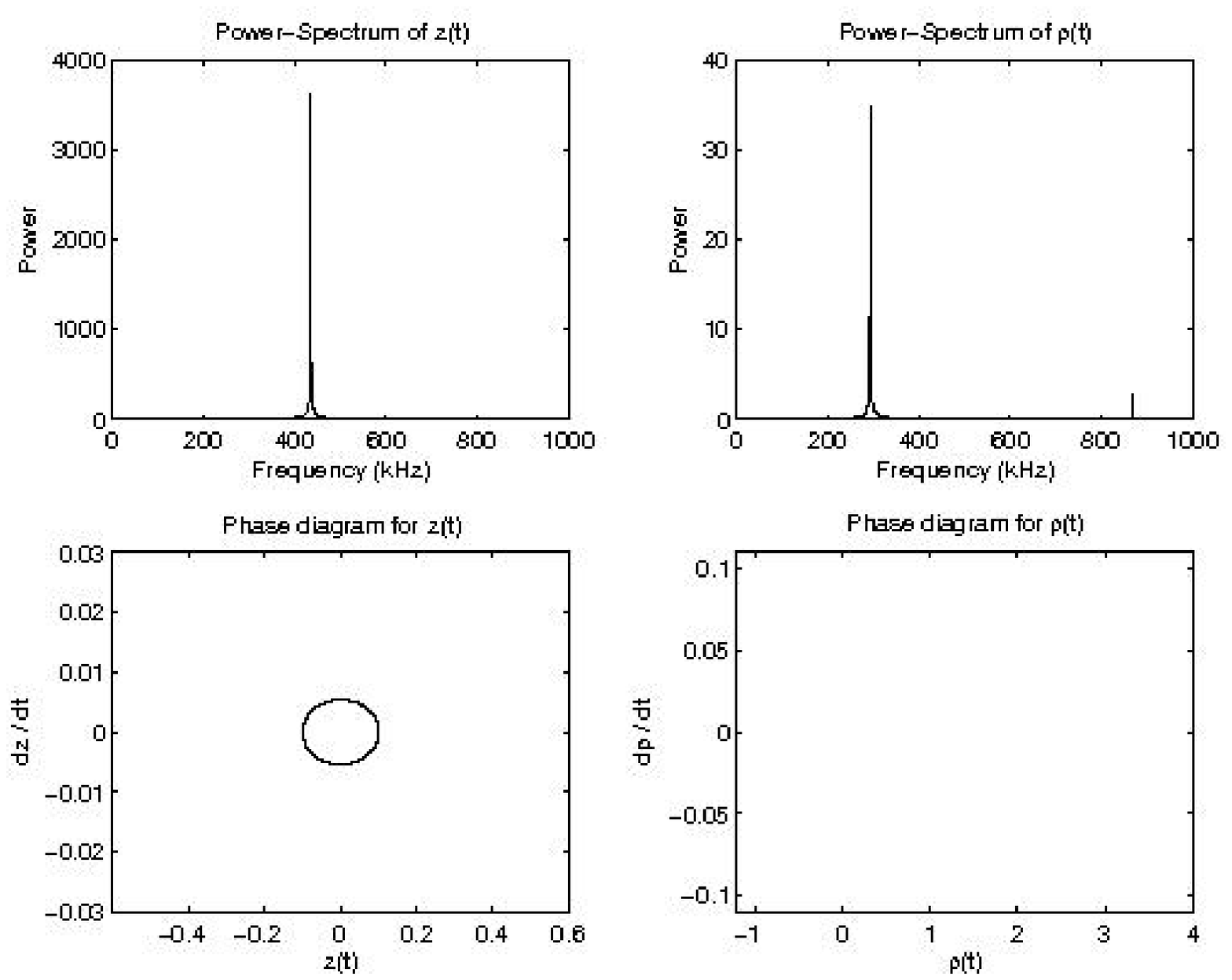}{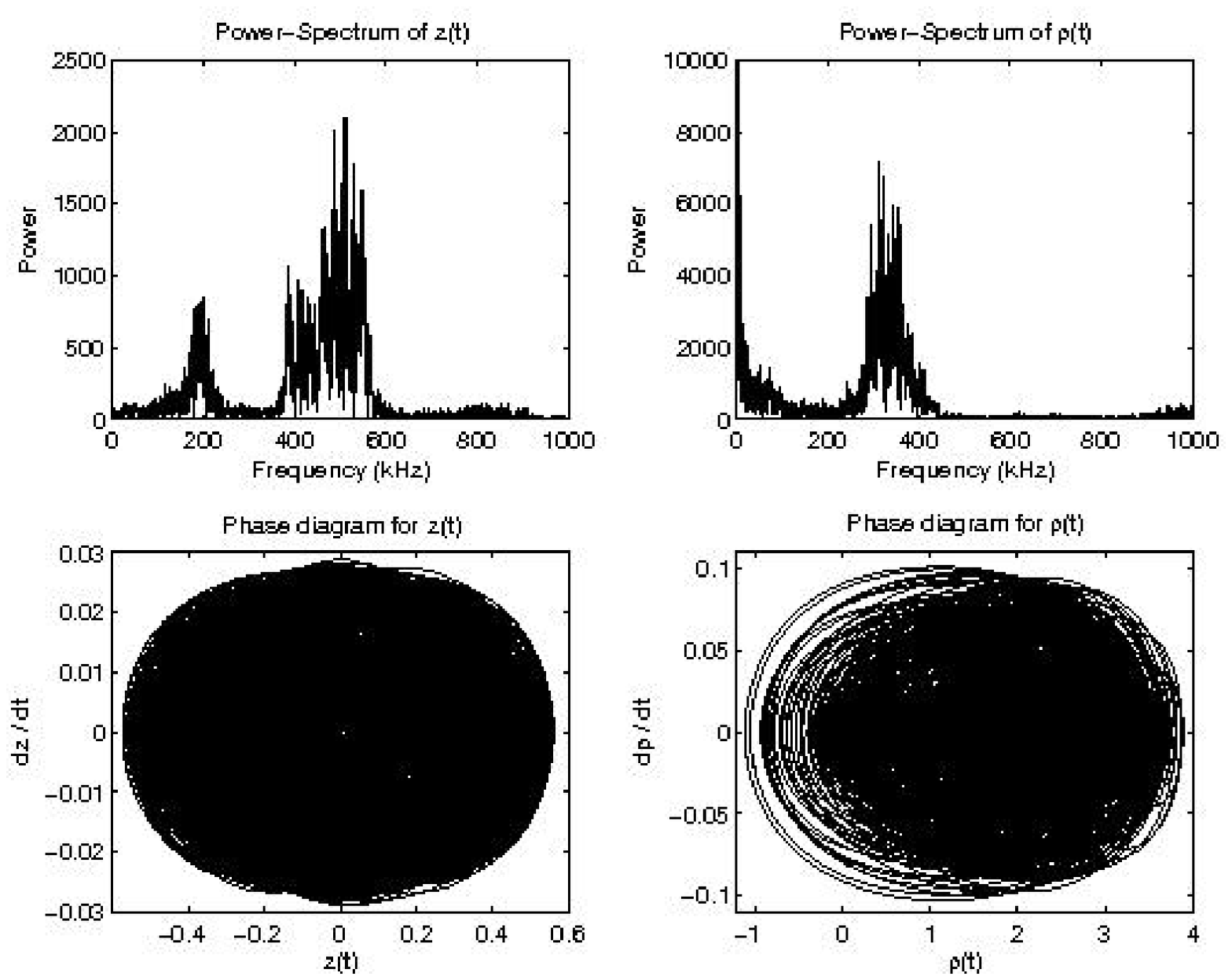}
\caption{Spectra (upper panels) and phase diagrams (lower panels) for radial ($\rho$) and vertical ($z$) perturbed geodesics. The displacements are in units of Schwarzschild radii, the frequencies are scaled to kHz (assuming a central mass $M$ of $2~M_\odot$). The left plots are noise free, while the right have a small noise in the $z$ direction. Clearly the amplitudes of oscillation are much greater in presence of stochastic forcing. Note that if turbulence becomes too strong, then any periodicity is lost and the motion becomes chaotic. See section \ref{subsec:bh} and the original paper \citep{vio06} for a more extensive explanation. }
\label{fig:turbo}
\end{figure*}

Four microquasars display pairs of HFQPOs, whose centroids ratio is consistent with $3:2$ \citep[][and see Figure \ref{fig:bursa}]{rem06}. Sometimes the twin peaks are observed simultaneously, most often only one is present.
GRS 1915+150 has a second pair \citep{str01} in a 5:3 frequency 
ratio \citep{klu02}.\\
 Contrarily to what it may seem at first sight, all these features are naturally explained in the resonance model. Indeed the absence of a peak may be due to the relativistic modulation of the emitted light \citep{bur04}.
Moreover although the strongest resonance is expected to occur when the frequencies are in $3:2$ ratio, lower order resonances (such as $5:3$) and subharmonics can still  be excited and become visible \citep{abr03,reb04,hor04}. In the resonance picture the vicinity of the observed ratio to commensurate ratio indicates that the frequency corrections are small. \\
The real problems of the model are more substantial: how are the QPOs excited?
 How are they coupled? In a preliminary toy-model based on the idea of \citet{abr01}, the motion of a single particle on a perturbed geodesic was analyzed. In this simple model
 an ad-hoc parameter was added \citep{abr03,reb04}, that would provide the energy to increase the amplitudes of the resonant oscillations. 
This approach works well mathematically, but the physics remains unexplained.  A more detailed fluid model of the disk should be developed in order to justify this excitation mechanism and to understand whether for example a form of disk instability \citep[e.g.,][or magnetic instability]{pap84} could arouse the oscillations.
\\
Accretion disks are  highly turbulent.
\citet{vio06} investigated the possibility that stochastic turbulence could excite and feed the resonance. To this purpose a white noise term was added to the vertical component of the perturbed geodesic and it was found that a small noise could indeed trigger the epicyclic resonance (see figure \ref{fig:turbo}). This result opens an interesting  perspective. Still a more realistic scenario, that takes into account colored noise and  stochastic forcing of a disk (rather than a single particle) should be evaluated.\\
Finally in the alpha of the QPOs resonance theory it was always mentioned that gravity and pressure would provide the coupling between the resonant modes. However it was recently estimated that in thin disks and slender tori such a coupling is too weak to actually transfer energy between epicyclic modes in a reasonably short time (Ji\v{r}\'i Hor\'{a}k, private communication). This difficulty holds in neutron star systems as well.
One possibility could be to consider the coupling of epicyclic modes in different geometric configurations. Still this may originate new dilemmas: for instance, in a thick torus the epicyclic oscillations (if not damped) do not simply scale with the mass of the central object, but depend strongly on thermodynamical quantities. Moreover \cite{sra07} showed that in a symmetric newtonian slender/thick torus there is no pressure coupling between epicyclic modes. Non-axisymmetry may play an important role.\\
Another option would be to consider non-epicyclic modes, but one should understand why two non-epicyclic modes are more important than others. 
Although much has been done, the excitation and coupling of modes still needs to be developed and a close interaction between analytic and numerical work is fundamental. Simulated axisymmetric tori slightly off equilibrium do not incontrovertibly produce QPOs : they need to be "kicked" \citep[][]{lee04,zan05,bla06}.   Once the modes are excited (via velocity, density perturbations or periodic forcing), then a spectrum of modes arise, with some modes in $3:2$ ratio. One of the challenges consists to identify the simulated modes with analytic solutions and to tag if there is any resonances.\\
We conclude this section on microquasars pointing at another puzzle:  \cite{tor06} reported that the spin estimates from the resonance model do not match the estimates obtained by fitting spectral continua.  This is a puzzle that any relativistic QPOs model must address.

\subsection{Difficulties in Neutron Star Systems}
\label{subsec:ns}

\begin{figure}
\centering \leavevmode
\epsfxsize=\columnwidth \epsfbox{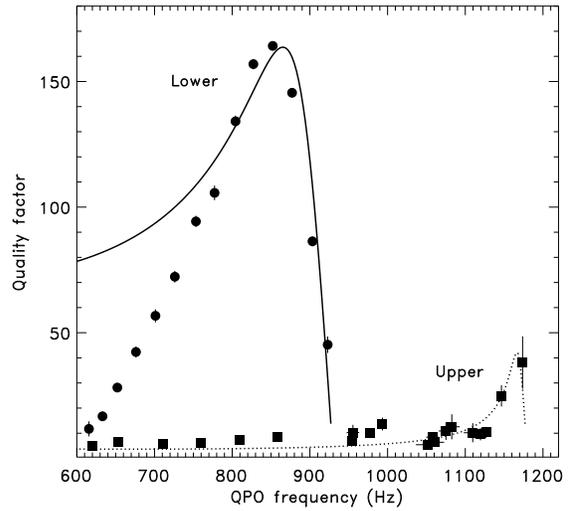}
\caption{Measured quality factor of the lower (dots) and upper (squares) HFQPOs of
 4U 1636-536. The solid  and dashed lines show the estimates done with the toy model
proposed in \cite{bar06}, from which this figure is taken. }
\label{fig:q}
\end{figure}

As we have seen there is no general agreement about the nature of HFQPOs in black hole systems. For neutron stars the situation is even more complex and the scientific community is torn between those who think that HFQPOs in neutron star systems have the same origin as in microquasars, and those who sustain that it is a different phenomenon.
In the scenario proposed by \cite{abr01} HFQPOs have the same root in both classes of sources, with some important caveats.\\
Twin HFQPOs are observed in accreting neutron stars: however both peaks can be in a different position at a different time \citep[e.g.,][]{van05}. The frequency shift can be higher than a factor of $2$ (see Figure \ref{fig:bursa}). The ratio of the two centroids is sometimes in the vicinity of $3:2$, other times it differs from the exact ratio (up to $\sim 10 \%$). \cite{bel05} argue that in neutron star sources there is not a preferred fixed  ratio \citep[as][claim]{abr03c} and the occurrence of ratios close to $3:2$ is biased. \\
In the resonance model these QPOs are still interpreted as resonant epicyclic modes, with amplitudes larger than in the case of black holes. The neutron star surface plays an important role: the X-ray emission is indeed modulated at the boundary layer, where the accreted matter hits the star  \citep{hor05,abr07}. The divergence from the exact ratio and the possibility of more than one ratios \citep[as suggested by][]{bel05} are natural consequences of the nonlinear nature of the oscillators \citep[e.g.,][]{lan69} \\
The bright Sco X-1 is often taken as a paradigm for neutron star sources. Different detections lie on a line with a slope close, but not equal to $3:2$ \citep{van97}. 
\cite{abr03} and \cite{reb04} reproduced part of this line, by considering the frequencies and ratio shifts as frequencies corrections ($\Delta \nu(t)$) due to nonlinear resonance. This result is qualitatively interesting, but the toy-model they adopt is too simple. Moreover in the nonlinear theory the lowest order frequency corrections scale with the square of the amplitude of the perturbation. Shifts as large as those observed would mean so high amplitudes that the adopted  expansion would fail.\\
HFQPOs in neutron star sources may be related to the spin of the central object.
In the very first detections the  difference of the frequencies of the two peaks was found to be consistent with being constant and equal to the spin of the neutron star. When more sources were observed, they were divided into two classes: slow ($\nu_{spin}<400$ Hz) and fast rotators, in which the difference was equal to the spin or half of it respectively \citep{mun01}.
From the study of Sco X-1 \cite{abr03b}  pointed that HFQPOs have "little to do with the rotation of a stellar surface or any magnetic field structure anchored in the star".
Very recently \cite{men07} re-examined all the available data and found that the difference of HFQPOs is approximately constant but not related to the spin and this issue is currently under debate \citep{str07,bar07}. Models of HFQPOs became more or less popular each time that a correlation or not with the spin was found \citep[see][for a review apropos]{men07}. The original structure of the resonance model is unrelated to the spin of the central object. It was however subsequently proposed \citep[][]{lee04} that in neutron star sources the resonance could be excited by direct forcing of the rotating star, leading to frequency differences equal to the spin or half of it. 
It is worth noticing that this possibility is not structural in the resonance model, but is a variation of it. On the other hand this brings back the problem of the excitation of the modes, like in the case of black hole systems.\\
Eventually another aspects remains to be explained, that is the behavior of the quality factor of HFQPOs ($Q = \nu / \text{FWHM}$), that accounts for the coherence of the oscillations. \cite{bar06} reported that the quality factor of the lower HFQPOs is higher than 
 that of the upper. Furthermore, while the $Q$ of the lower frequencies increases with the frequency up to a break frequency and then drops, the $Q$ of the upper frequencies increases steadily (see Figure \ref{fig:q}). These measurements suggest that in different sources there is a common physical mechanism not only to excite the oscillations, but also to damp them. 
Hopefully this conduct of the quality factor may help the theory to to shed some light on these mechanism, that for the moment remain obscure.

\section{Conclusions}
\label{sec:concl}

The nonlinear resonance model for high frequency QPOs encounters many internal difficulties. This may mark the end of the model or the beginning of a new era.\\
In our opinion the presence of twin peaks in (almost) commensurate ratio and the detection of subharmonics still strongly support the idea that weak nonlinear resonance is a fundamental ingredient in the physics of HFQPOs.\\
It is like smelling chocolate in a birthday cake: the smell is the unmistakable signature of the presence of chocolate. However the other (mathematical and physical) ingredients, their relative importance and a draft of the recipe (mechanisms) are needed to know  which cake it is.

% \bibitem[Names(Year)]{label} or \bibitem[Names(Year)Long names]{label}.
% (\harvarditem{Name}{Year}{label} is also supported.)

\bibliographystyle{harvard}
\bibliography{biblio}

%\begin{thebibliography}{}
%\bibitem[]{}
%\end{thebibliography}

\end{document}